%% file: WRCMpaperfinal.tex
\documentclass[]{tWRM2e}
\input macro
\begin{document}
\doi{10.1080/1745503YYxxxxxxxx}
 \issn{1745-5049}
\issnp{0278-1077}
\jvol{00} \jnum{00} \jyear{2008} \jmonth{January}

\markboth{Taylor \& Francis and I.T. Consultant}{Waves in Random and Complex Media}

%\articletype{paper}

\title{Symmetry and resonant modes in platonic grating stacks}

\author{S. G. Haslinger$^{\rm a}$$^{\ast}$\thanks{$^\ast$Corresponding author. Email: sgh@liv.ac.uk
\vspace{6pt}}, A. B. Movchan$^{\rm a}$, N. V. Movchan$^{\rm a}$ and R. C. McPhedran$^{\rm a,b}$\\\vspace{6pt}  $^{\rm a}${\em{Department of Mathematical Sciences,
Mathematical Sciences Building, 
 Peach Street, Liverpool L69 3BX,
United Kingdom}}; $^{\rm b}${\em{CUDOS, School of Physics, University of Sydney 2006, NSW, Australia}}
\\\vspace{6pt}\received{received May 2013} }

\maketitle

\begin{abstract}
\bigskip

We study the flexural wave modes existing in finite stacks of gratings containing rigid, zero-radius pins. We group the modes into 
even and odd classes, and derive dispersion equations for each. We study the recently discovered EDIT (elasto-dynamically inhibited transmission) phenomenon, and relate it to the occurrence of trapped waves of even and odd symmetries being simultaneously resonant. We show how the EDIT interaction may be steered over a wide range of frequencies and angles, using a strategy in which the single-grating reflectance is kept high, so enabling the quality factors of the even and odd resonances to be kept large.

\bigskip

\end{abstract}

\section{Introduction}
The topic of this paper is the interaction of flexural waves with finite stacks of gratings consisting of fixed pins. The flexural waves satisfy the biharmonic equation, and may be thought of as superpositions of a part obeying the Helmholtz equation, and a part obeying the modified Helmholtz equation. This superposition in fact guarantees that the flexural waves have interesting and unusual properties - for example, their Green's function is finite at the source point, rather than diverging logarithmically there as is the case for the two-dimensional Green's function of the Helmholtz equation.
These properties have been explored in a series of recent papers from several groups [1-15]. Among striking results from these papers, we note that the square lattice of pinned points exhibits a total band gap at low frequencies, and, as distinct from the case of the Helmholtz equation, the gap width does not tend to zero as the pin radius tends to zero. This means that the flexural wave properties
of gratings made of fixed pins having zero radius are non-trivial, and such gratings when stacked together can give good filtering action, an effect that can be accurately described using a simple theory.
This makes the interaction of flexural waves with stacked gratings of zero-radius pins a useful problem in wave theory, with powerful and practically interesting results coming from an easily-understood
model.

In a recent paper \cite{has1} we identified a filtering effect  similar to  the quantum atomic effect called Electromagnetically Induced Transparency (EIT). The Elasto-Dynamically Inhibited Transmission (EDIT) is a classical physics effect characterised by a resonant maximum in transmission being cut in two by a resonant minimum with an extremely high quality factor. It is dependent on the coincidence of 
even and odd modes (where even and odd refer to parity in the $y-$coordinate, measured from the central grating).  For a triplet consisting only of rigid pins, the 
odd mode's frequency is invariant with the relative lateral shift  of a central grating   \cite{has1} . This symmetry breaking layer makes it easier to tune the system's 
even mode to coincide with the 
odd mode. This is harder but still possible for inclusions of nonzero radius.
Haslinger {\it et. al} \cite{has2} discuss how the radius of the inclusions affects EDIT, and give examples  for EDIT within structures containing gratings of inclusions of nonzero radius. 
The increased radius makes it necessary to take into account higher-order multipole terms characterising the scattered field, and the periodicity of the grating leads to the use of higher order grating sums. It follows that more care is required to identify the parameter values for EDIT for structures incorporating finite-sized voids. A limitation of both these papers is that the EDIT phenomenon is limited to a narrow range of frequency and angle of incidence values. It is a principal aim here to remove these restrictions, and show how the EDIT phenomenon can be steered over a wide range of these two parameters.

In Section 2, we derive the plane wave expansion of the Green's function for the biharmonic equation. This will be employed in later sections, replacing the cylindrical function expansion which we have employed in our previous papers on the EDIT phenomenon. In Section 3, we consider stacks of three gratings of rigid pins, with the central grating not necessarily aligned with the outer pair of gratings.
We construct the dispersion equations for modes in the triplet, considering both modes which are antisymmetric and symmetric as functions of the coordinate $y$ (see Fig. 1) orthogonal to the gratings.
We also investigate the conditions for odd and even modes to coincide.  
In Section 4, we concentrate on triplets in which all three gratings are aligned. In Section 5, we describe the procedure we have developed which enables us to steer modes in general and the EDIT phenomenon in particular  over a wide range of angles and frequencies. The method involves choosing (say) the angle of incidence, then finding the frequency parameter which delivers a  single-grating reflectance as close as desired to unity. The grating separation in the triplet stack is then chosen to make the frequency parameter correspond to a resonance of the odd mode  in the triplet. The final step is to shift the central symmetry-breaking layer (SBL) to move the frequency of the even mode to the desired value. We demonstrate the effectiveness of this strategy for angles of incidence ranging from $1^\circ$ to $60^\circ$, for which the biharmonic frequency parameter $\beta$ varies from 4.454 down to 2.947.

A preliminary account of some of the results presented here has been given in reference [16].

\section{Grating Green's function - plane wave form}
We consider a single grating of rigid pins as a line of point forces with constant separation $d$. Therefore we use a quasi-periodic Green's function $G( x, y; \alpha_0, \beta)$ for the biharmonic operator, satisfying the equation
\begin{equation}
( \Delta^2 - \beta^4) G (x,y; \alpha_0, \beta) \, + \, \delta (y) \sum_{n = -\infty}^{\infty} \, \delta (x - nd) \exp\{i \alpha_0 \, nd\} \, = \, 0,
\end{equation}
where $\alpha_0$ is the Bloch parameter and $\beta$ is the spectral parameter associated with the frequency $\omega$ by $\beta^2 = \omega \sqrt{\rho h/D}$. In this paper it is sufficient to use the plane wave form, rather than the spatial form which incorporates Bessel functions and grating sums. The spectral form of the Green's function is
\begin{equation}
G(x, y; \alpha_0, \beta)  = -\frac{1}{2 \beta^2} \bigg( \frac{1}{2 i d} \sum_{n = -\infty}^{\infty}  \frac{1}{\chi_n} e^{i[\alpha_n x \, + \,  \chi_n |y|]}  +  \frac{1}{2d} \sum_{n = -\infty}^{\infty} \frac{1}{\tau_n} e^{i \alpha_n x} \, e^{- \tau_n |y|} \bigg), 
\end{equation}
where 
\begin{eqnarray}
\alpha_n & = &  \alpha_0  +  \frac{2 \pi n}{d},    \label{alph1}\\ 
\chi_n & = & \bigg\{ \begin{array} {l l} \sqrt{\beta^2 \, - \, {\alpha_n}^2}  & , \, \, {\alpha_n}^2 \le \beta^2, \\  i \sqrt{ {\alpha_n}^2 \, - \,\beta^2} & , \, \, {\alpha_n}^2 > \beta^2, 
 \end{array}  \label{alph2}  \\
 \tau_n & = & \sqrt{\beta^2 + \alpha_n^2}    \label{alph3} .
\end{eqnarray}

As mentioned by Evans and Porter (2007), the Green's function for the biharmonic operator is cubically convergent, given that the two sums are considered together. We also note that it is necessary to use a far greater number of terms for evaluations of the Green's function at the origin than for other gratings within a stack (typically 1000 and 20 respectively).
\begin{figure}
\begin{center}
\subfigure[]{
\resizebox*{6.5cm}{!}{\includegraphics{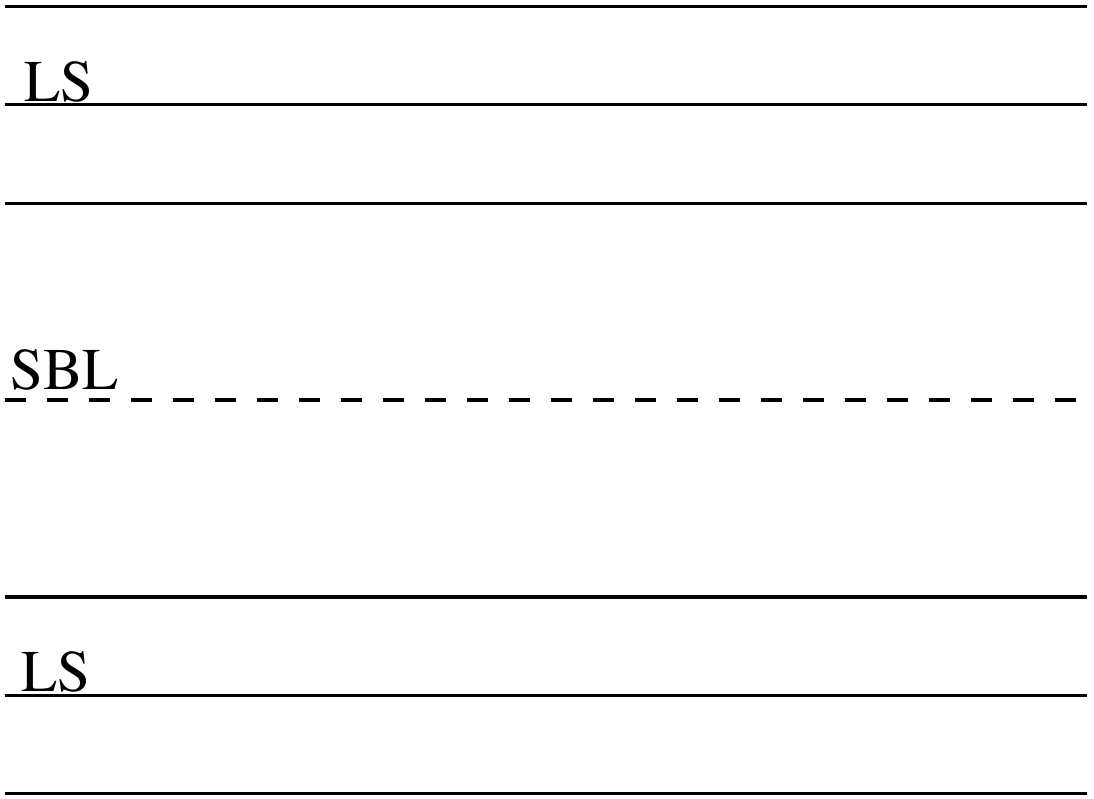}}}
\subfigure[]{
\resizebox*{6.5cm}{!}{\includegraphics{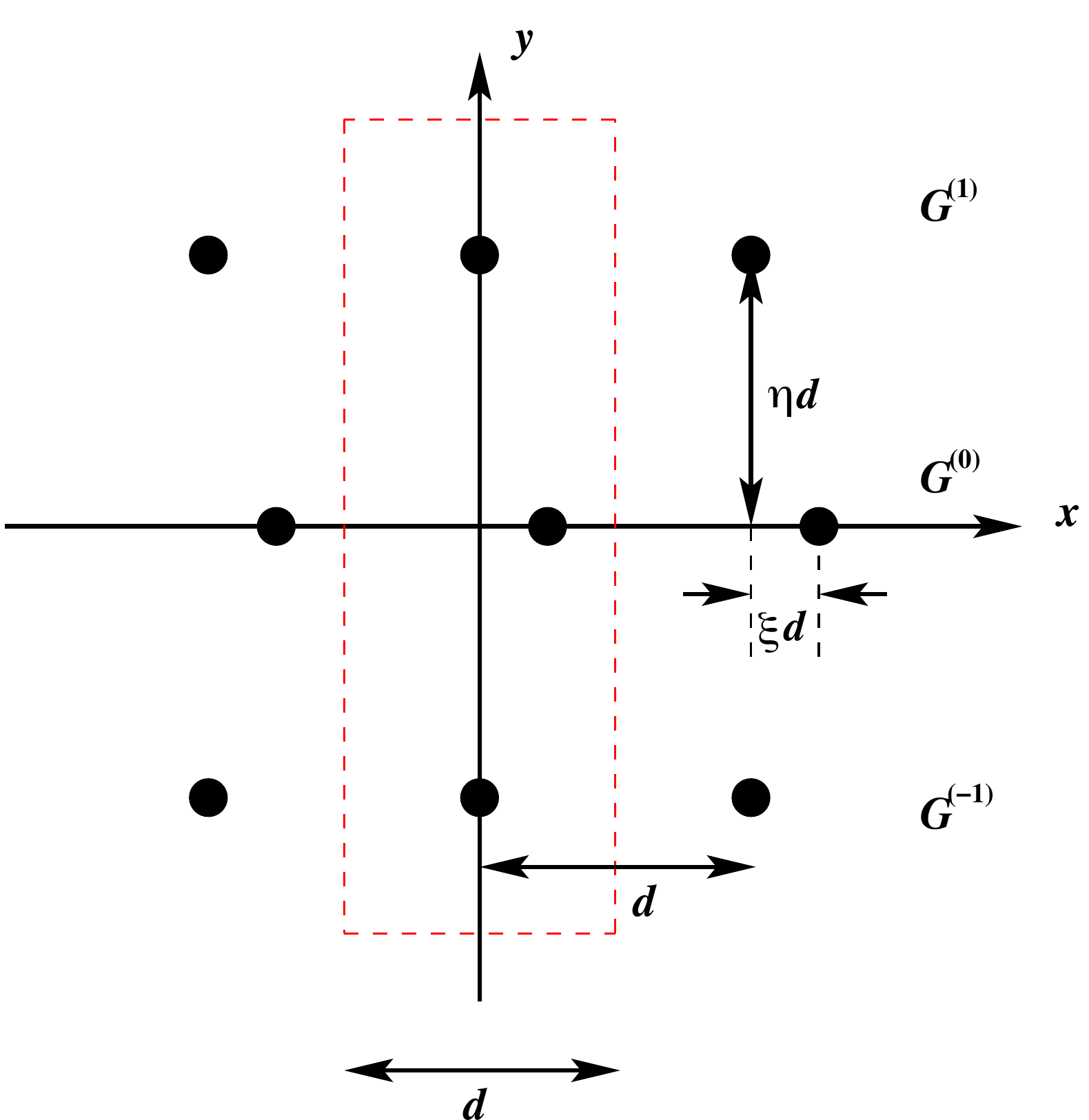}}}
\caption{Example of a general grating waveguide.}
\label{genstruct}
\end{center}
\end{figure}

Our general structure consists of two layer stacks (LS) and a symmetry breaking layer (SBL) as shown by Fig.~\ref{genstruct}(a). The identical layer stacks contain a finite number of periodic gratings $N$ and act as symmetric mirrors. Note that with the mirror systems chosen to preserve up-down symmetry, there is always an odd number of layers. Of course the simplest example is when $N = 1$, producing a triplet, as shown in Fig.~\ref{genstruct}(b).

\section{Mode symmetry in a general triplet stack}
\label{sect3}

The elementary cell for the general triplet is illustrated in Fig.~\ref{genstruct}(b). The period of each grating is $d$ and the relative vertical separation between gratings is $\eta$. The relative lateral shift of the central grating (or SBL) is denoted by $\xi$.
For the flexural displacement $u$
, we write the sum of three quasi-periodic Green's functions:
\begin{equation}
u \, = \, \sum_{j = -1}^1 \, A_j \, G^{(j)} \, (x, \,y ; \, \alpha_0, \, \beta) , \label{QPU}
\end{equation}
where $A_j$ are coefficients to be determined for each Green's function defined for the index $j$. At each pin, it is required that $u = 0$. We therefore take the elementary column cell incorporating the central pin of each of the constituent gratings such that
\begin{equation}
u \Bigg|_{{\bf x} = \scriptsize{\textbf{\emph{a}}^{(m)}}} 
\, = \, 0,   \,\,\,\,\,\,\,\,\,\, \mbox{where} \,\,\,\, m = -1, 0, 1.
\end{equation}
For a given choice of $\textbf{\emph{a}}^{(m)}$, we require that 
\begin{equation}
u \Bigg|_{{\bf x} = \scriptsize{\textbf{\emph{a}}^{(m)}}} \, = \, \sum_{j = -1}^1 \, A_j \, G^{(j)} \, ({\textbf{\emph{a}}^{(m)}}
; \, \alpha_0, \, \beta) \, = \, 0,  \label{meq}
\end{equation}
which is equivalent to the matrix equation 
\beq
\BM \, \BA \, = {\bf 0}.
\eequ{MA}
Here ${\textbf{\emph{a}}^{(0)}} = (\xi d, 0)$ and ${\textbf{\emph{a}}^{(\pm 1)}} = (0, \pm \eta d)$.

We consider four grating Green's functions; $G^{(1)}$, $G^{(0)}$ and $G^{(-1)}$ for the shifted waveguide, and a canonical Green's function, without an index, which we use to represent all matrix elements, $G(x, y)$ (see Fig.~\ref{genstruct}(b)). This Green's function is evaluated for the points $(n d, 0)$ on the horizontal axis.

We want to evaluate the quasi-periodic Green's function at $(0, \eta d)$, $(\xi d, 0)$ and $(0, -\eta d)$, i.e. at the pins within the elementary cell of the shifted waveguide:

\begin{equation}
\hspace{-0.1in}(0, \eta d):  
A_1 G^{(1)} (0, \eta d; \alpha_0, \beta)  + 
A_0 G^{(0)} (0, \eta d; \alpha_0, \beta)  + 
A_{-1} G^{(-1)} (0, \eta d; \alpha_0, \beta)  =  0,
\label{g1}
\end{equation}
\begin{equation}
\hspace{-0.1in}(\xi d, 0): 
A_1 G^{(1)} (\xi d, 0; \alpha_0, \beta)  + 
A_0 G^{(0)} (\xi d, 0; \alpha_0, \beta)  + 
A_{-1} G^{(-1)} (\xi d, 0; \alpha_0, \beta)  = 0,
\end{equation}
$$
\hspace{-2.0in}(0, -\eta d): 
A_1 G^{(1)} (0, -\eta d; \alpha_0, \beta)  + 
A_0 G^{(0)} (0, -\eta d; \alpha_0, \beta)  $$
\begin{equation}
+A_{-1} G^{(-1)} (0, -\eta d; \alpha_0, \beta)  =  0.
\end{equation}
In terms of the canonical Green's function $G(x, y)$ for the central grating,
\begin{equation}
\begin{array} {l l l} G^{(1)}(x,y; \alpha_0, \beta) & =  & G(x, y - \eta d; \alpha_0, \beta), \\ \\  G^{(0)}(x,y; \alpha_0, \beta) & =  & G(x - \xi d, y; \alpha_0, \beta), \\ \\ G^{(-1)}(x,y; \alpha_0, \beta) & = & G(x, y + \eta d; \alpha_0, \beta). \end{array}
\label{gc}
\end{equation}
Matrix $\BM$ is then filled according to equations~(\ref{g1})-(\ref{gc}):
\begin{equation}
{\BM}=\left( \begin{array}{ccc}
G{(0, 0)} & G{(-\xi d, \eta d)}  & G{(0, 2 \eta d)} \\ \\
G{(\xi d, -\eta d)} & G{(0, 0)}   & G{(\xi d, \eta d)}\\ \\
G{(0, -2 \eta d)}& G{(-\xi d, -\eta d)} & G{(0, 0)} \end{array} \right) \, = \, \left( \begin{array}{ccc}
M_{11} & M_{12} & M_{13} \\
M_{21} &M_{11}  & M_{21}\\
M_{13} & M_{12} & M_{11}\end{array} \right),
\end{equation}
using $G(x, -y) = G(x, y)$, and where the arguments $\alpha_0$ and $\beta$ in the representation of the matrix elements $M_{ij}$ are left implicit.

Note that the elements $M_{12}$ and $M_{21}$ are in general unequal if the central layer is shifted with respect to the upper and lower layers. An exception to this arises for normal incidence when $\alpha_0 = 0$. From equations (\ref{alph1})-(\ref{alph3}) we see that when $\alpha_0 = 0$ then $\alpha_{-n} = -\alpha_n$, $\chi_{-n} = \chi_n$, and $\tau_{-n} =\tau_n$. It then follows that 
$$
M_{12} = G(-\xi d,\eta d) = M_{21} = G(\xi d, - \eta d).
$$

If the gratings are all aligned ($\xi = 0$), or we deal with the case $\alpha_0=0$ the matrix $\BM$ has a symmetric Toeplitz structure. This would be the case regardless of the number of gratings chosen for the mirror layers placed symmetrically above and below the SBL.
In the analysis below we focus on the three-gratings stack. 

The matrix $\bf M$ has three eigenvalues:
\begin{equation}
 \lambda_1=M_{11}-M_{13},~~ \lambda_\pm=\frac{1}{2}(2 M_{11}+M_{13}\pm \sqrt{8 M_{12}M_{21} + M_{13}^2}).
\end{equation}
The eigenvector corresponding to $\lambda_1$ is
\begin{equation}
\label{oddev}
{\boldmath v}_o=\left[\begin{array}{r} -1\\0\\1\end{array}\right],
\end{equation}
and has odd symmetry. The other two vectors have even symmetry:

\begin{equation}
{\boldmath v}_{e-}=\left[\begin{array}{c} 1\\
{\left(- M_{13} - \sqrt{8 M_{12}M_{21} + M_{13}^2}\right)}/{(2 M_{12})}\\1\end{array}\right], \label{vm}
\end{equation}
and
\begin{equation}
{\boldmath v}_{e+}=\left[\begin{array}{c} 1\\
{\left(- M_{13} + \sqrt{8 M_{12}M_{21} + M_{13}^2}\right)}/{(2 M_{12})}\\1\end{array}\right]. \label{vp}
\end{equation}

The dispersion curves for modes are trajectories along which the eigenvalues are zero. Thus for the odd mode, the dispersion curve corresponds to the condition:
\begin{equation}
\label{oddc}
M_{11}-M_{13}=0 .\end{equation}
Along this trajectory, the other two eigenvalues of $\BM$ are
\begin{equation}
\lambda_\pm=\frac{1}{2}\left(3M_{13}\pm\sqrt{8 M_{12}M_{21} + M_{13}^2}\right),
\end{equation}
with eigenvectors as in \eq{vm}, \eq{vp}.

For the even modes, the dispersion curves correspond to 
\begin{equation}
\label{evenc}
2 M_{11}+M_{13}=\mp \sqrt{ 8M_{12}M_{21} + M_{13}^2}.
\end{equation}
Alternatively,
$$
M_{12} = \frac{M_{11} (M_{11} + M_{13})}{2M_{21}} ~{\rm or}~ M_{13} = -M_{11} + \frac{2M_{12}M_{21}}{M_{11}}.
$$
The even eigenvectors for the shifted waveguide are not orthogonal, whereas they are for the unshifted case.
Replacing $M_{12}$ 
according to the above condition required for the even mode's dispersion curves, we obtain the matrix
\begin{equation}
\BM=\left( \begin{array}{ccc}
M_{11} & \frac{M_{11} (M_{11} + M_{13})}{M_{21}} & M_{13}\\ \\
M_{21} &M_{11}  & M_{21}\\ \\
M_{13} & \frac{M_{11} (M_{11} + M_{13})}{M_{21}}& M_{11}\end{array} \right),
\end{equation}
which has eigenvalues of 0, $M_{11} - M_{13}$ and $2M_{11} + M_{13}$, with corresponding eigenvectors:
\begin{equation}
\left[\begin{array}{c} 1\\
-\frac{2M_{21}}{M_{11}}\\1\end{array}\right] ~,~\left[\begin{array}{r} -1\\
0\\ 1\end{array}\right]~,~\left[\begin{array}{c}  1\\
\frac{2M_{21}}{M_{11}+M_{13}}\\ 1\end{array}\right].
\end{equation}

\subsection{Condition for even modes to coincide}
The condition for the two even modes to coincide is
 $$
 2 M_{11}+M_{13} - \sqrt{8 M_{12}M_{21} + M_{13}^2}
 = 
 2 M_{11}+M_{13} + \sqrt{8 M_{12}M_{21} + M_{13}^2}
 = 0.
 $$
 This gives us the conditions
 \begin{equation}
 M_{13} = -2M_{11} ~{\rm and \,\,} M_{21} = -\frac{M_{11}^2}{2M_{12}},
 \end{equation}
 from which we obtain the matrix
 \begin{equation}
\BM=\left( \begin{array}{ccc}
M_{11} & M_{12} & -2M_{11}\\ \\
-\frac{M_{11}^2}{2M_{12}} &M_{11}  & -\frac{M_{11}^2}{2M_{12}} \\ \\
-2M_{11} & M_{12} & M_{11}\end{array} \right),
\end{equation}
which has eigenvalues 0, 0 and $3M_{11}$. The matrix is defective and has two eigenvectors:
\begin{equation}
\left[\begin{array}{c} 1\\
\frac{M_{11}}{M_{12}}\\ 1\end{array}\right] ~,~\left[\begin{array}{r} -1\\
0\\ 1\end{array}\right].
\end{equation}
 This shows that it is impossible for the two even modes of the triplet to have an avoided 
 crossing, since for this to occur the null space of the matrix $\bf M$ at the crossing
 would have to have dimension two. 

\subsection{Condition for even and odd mode to coincide}
For the odd mode dispersion curve, we have the condition $M_{11} - M_{13} = 0$ and the matrix
\begin{equation}
\BM=\left( \begin{array}{ccc}
M_{11} & M_{12} & M_{11} \\
M_{21} &M_{11}  & M_{21}\\
M_{11} & M_{12} & M_{11}\end{array} \right).
\end{equation}
For an even mode's dispersion curve, we have $M_{12} = M_{11} (M_{11} + M_{13})/2 M_{21}$ so for the coincidence of an odd and an even mode, $M_{12} = M_{11}^2/M_{21}$ and 
\begin{equation}
\BM=\left( \begin{array}{ccc}
M_{11} & M_{12} & M_{11} \\ \\
\frac{M_{11}^2}{M_{12}} &M_{11}  & \frac{M_{11}^2}{M_{12}}\\ \\
M_{11} & M_{12} & M_{11}\end{array} \right),
\end{equation}
which has eigenvalues
0, 0 and $3M_{11}$, with corresponding eigenvectors
\begin{equation}
\left[\begin{array}{r}  -1\\
0\\ 1\end{array}\right]~,\left[\begin{array}{c}-\frac{M_{12}}{M_{11}}\\
 1\\0\end{array}\right] ~,~\left[\begin{array}{c}  1\\
\frac{M_{11}}{M_{12}}\\ 1\end{array}\right].
\end{equation}
The coincidence of the other even mode and the odd mode occurs with eigenvalues 0, 0 and $-3M_{11}$ with the corresponding eigenvectors
\begin{equation}
\left[\begin{array}{r}  -1\\
0\\ 1\end{array}\right]~,\left[\begin{array}{c}\frac{M_{12}}{M_{11}}\\
 1\\0\end{array}\right] ~,~\left[\begin{array}{c}  1\\
-\frac{M_{11}}{M_{12}}\\ 1\end{array}\right].
\end{equation}

\begin{figure}
\begin{center}
\includegraphics[width=6.8cm]{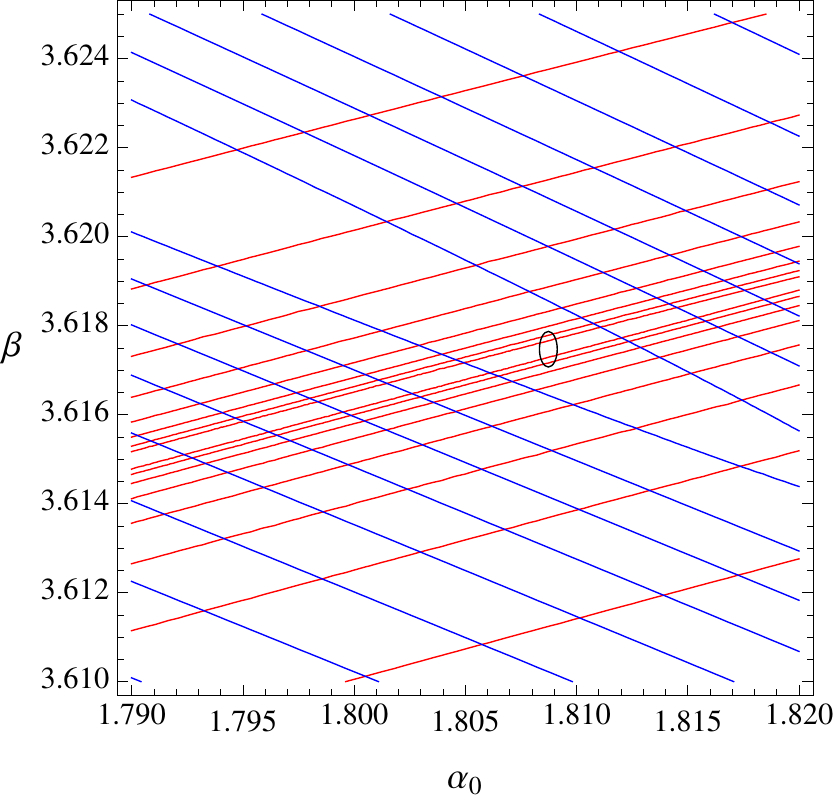}
\parbox{14cm}{\caption{\label{oec} Trajectories of constant  modulus of $M_{11}-M_{13}$ (in red), and of $2 M_{12}M_{21} - M_{11}(M_{11}+M_{13})$ (in blue) for a shifted three grating stack of rigid 
pins ($\xi = 0.25200d$),
as a function of $\alpha_0$ and $\beta$. The circle corresponds to $(\alpha_0,\beta)=(1.808735, 3.61747)$.}}
\end{center}
\end{figure}

As an example, we show in Fig.~\ref{oec} the trajectories corresponding to the odd mode (in red) and an even mode (in blue) for a stack of three rigid pin gratings that supports Elasto-Dynamically Inhibited Transmission (EDIT). The mode trajectories show the logarithm of the absolute value of the conditions~(\ref{oddc}) and~(\ref{evenc}), for which we seek zeros. Therefore the minimum modulus indicates the position of each mode's dispersion curve. The mode trajectories intersect in the vicinity of a point identified from reflection and transmission properties of the triplet as ($\alpha_0$, $\beta$) $ = (1.808735, 3.61747) $, an example of EDIT for $\theta_i = 30^{\circ}$ with lateral shift $\xi = 0.25200d$ of the central grating. 

The blue trajectories of minimum modulus (-14.4) are those that bracket one another in the central region of the space; the red trajectories of minimum modulus (-12.5) are the innermost red lines that concentrate around a bare region. The bare rhomboid region bounded by these four lines contains the EDIT point $(\alpha_0,\beta)=(1.808735, 3.61747)$, and the matrix $\BM$ from which we obtain the data, satisfies the conditions~(\ref{oddc}) and~(\ref{evenc}) to a degree of accuracy of $10^{-6}$ or smaller. 
We note that we have not been successful in locating examples of two even modes coinciding.

One effect it is useful to understand is the behaviour of matrix elements near light lines, where the matrix elements diverge. In fact, the elements contain terms which go as
\begin{equation}
-\frac{1}{4 \beta^2}\frac{e^{i \chi_m|p-n|d}}{i \chi_m},~ {\rm where}~ \chi_m = \sqrt{\beta^2 - \bigg(\alpha_0 +\frac{2\pi m}{d}\bigg)^2}.
\label{munsh14}
\end{equation}
At the light lines, the  $\chi_m$ go to zero for a particular  value of $m$, being real if $\beta$ is below the light line, and imaginary if it is above it. This will then give either a divergent term in the real part or the imaginary part of every element of $\BM$. Putting $X_m=e^{i \chi_m d}$, we find that the matrix $\BM$ contains a scaling term which goes as $-1/(4\beta^2 i \chi_m)$ times the light line matrix:
\begin{equation}
\BM_{LL}=\left( \begin{array}{ccc}
1 & X_m & X_m^2 \\
X_m &1  & X_m\\
X_m^2  & X_m & 1\end{array} \right).
\label{munsh15}
\end{equation}
This matrix has respective eigenvalues and eigenvectors
\begin{equation}
1-X_m^2, ~\frac{1}{2}(2+X_m^2-X_m\sqrt{8+X_m^2}),~ \frac{1}{2}(2+X_m^2+X_m\sqrt{8+X_m^2}),
\label{munsh16}
\end{equation}
\begin{equation}
\left[\begin{array}{r} -1\\
0\\ 1\end{array}\right],~~\left[\begin{array}{c}  1\\
\frac{1}{2}\left(-X_m-\sqrt{8+X_m^2} \right)\\ 1\end{array}\right],~~\left[\begin{array}{c}  1\\
\frac{1}{2}\left(-X_m+\sqrt{8+X_m^2} \right)\\ 1\end{array}\right].
\label{munsh17}
\end{equation}
Exactly on the light line, the eigenvalues tend to $0,0,3$ and the eigenvectors tend to
\begin{equation}
\frac{1}{\sqrt{2}}\left[\begin{array}{r} -1\\
0\\ 1\end{array}\right],~~\frac{1}{\sqrt{6}}\left[\begin{array}{r}  1\\
-2\\ 1\end{array}\right],~~\frac{1}{\sqrt{3}}\left[\begin{array}{c}  1\\
 1\\ 1\end{array}\right].
\label{munsh18}
\end{equation}
It will be noted that these eigenvectors have been written in symmetrised form.

\subsection{Transmission problem and an inhomogeneous algebraic system}

For the case when there is a wave incident on the grating stack, equation \eq{MA} for the coefficients in the representation \eq{QPU} of the displacement field can be replaced by
\beq
\BM \BA = -\BU^{(incident)},
\eequ{MA1}
where $\BU^{(incident)}$ is the vector of values of the incident field at the nodes $\Ba^{(j)}$ marked in the grating centre points in Fig. \ref{genstruct}(b).

This linear system can be solved to find the coefficients $A_{\pm 1}, A_0$ and hence evaluate the transmitted and reflected energies.
A corresponding analysis of the transmission problem has been published in \cite{has1}, and in the text below we refer to that paper when quoting the results regarding the transmission resonances.

\section{Mode symmetry in an unshifted triplet stack}
We consider the modes of the set of three gratings of rigid pins, with the gratings all aligned, leading to a structure with up-down and left-right symmetry. The Green's function matrix in this case
is complex and symmetric:
\begin{equation}
\BM=\left( \begin{array}{ccc}
M_{11} & M_{12} & M_{13} \\
M_{12} &M_{11}  & M_{12}\\
M_{13} & M_{12} & M_{11}\end{array} \right).
\label{munsh1}
\end{equation}

The results of Section~\ref{sect3} can be used for this special case, with $M_{12}M_{21}$ being replaced by $M_{12}^2$. The crossing of the odd mode trajectory with that of the even mode occurs when $M_{11} = M_{13}$, $M_{12}=M_{11}$, and for the eigenvalues $0$, $0$, $3 M_{11}$ the eigenvectors are

\begin{equation}
\frac{1}{\sqrt{6}}\left[\begin{array}{r} 1\\
-2\\ 1\end{array}\right]~,~\frac{1}{\sqrt{2}}\left[\begin{array}{r} -1\\
0\\ 1\end{array}\right] ~,~\frac{1}{\sqrt{3}}\left[\begin{array}{r}  1\\
 1\\ 1\end{array}\right].
\label{munsh12}
\end{equation}
For the second mode trajectory, the eigenvalues are the same, and the eigenvectors are
\begin{equation}
\frac{1}{\sqrt{6}}\left[\begin{array}{r}  1\\
 2\\ 1\end{array}\right] ~,~ \frac{1}{\sqrt{2}}\left[\begin{array}{r} -1\\
0\\ 1\end{array}\right] ~,~\frac{1}{\sqrt{3}}\left[\begin{array}{r}  1\\
-1\\ 1 \end{array}\right].
\label{munsh13}
\end{equation}

\begin{figure}
 \begin{center}
\includegraphics[width=3.0in]{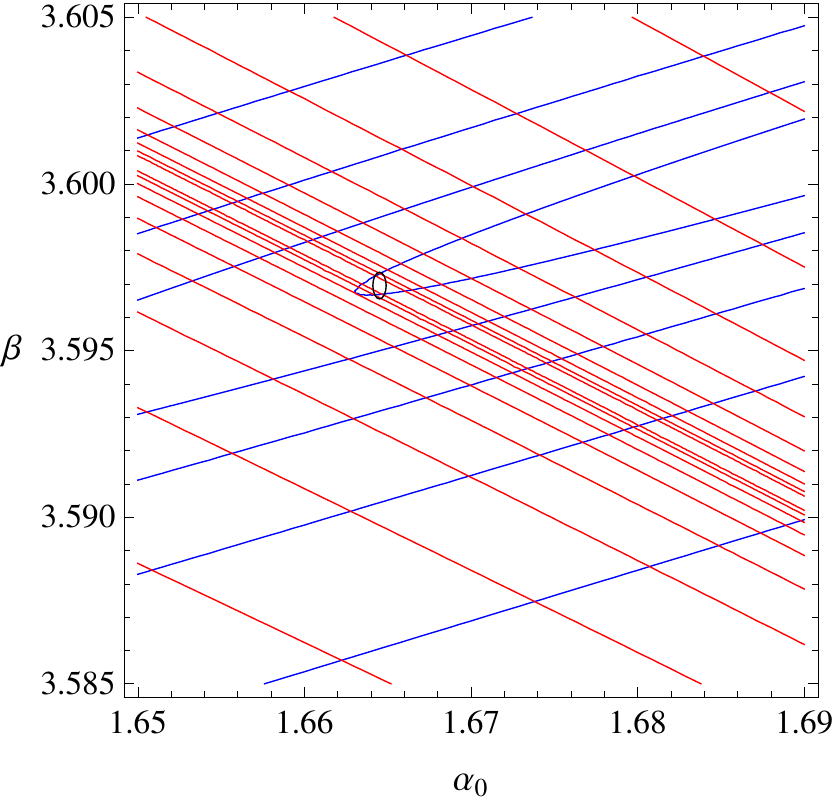}
\caption{\label{fig13ex}   Trajectories of constant  modulus of $M_{11}-M_{13}$ in blue, and of $2 M_{12}^2= M_{11}(M_{11}+M_{13})$ in red for an aligned three grating stack of rigid pins,
as a function of $\alpha_0$ and $\beta$. The circle corresponds to $(\alpha_0,\beta)=(1.66451, 3.596951)$.}
\end{center}
\end{figure}

As an example for the unshifted triplet, we show in Fig. \ref{fig13ex} trajectories corresponding to the odd mode (in blue) and an even mode (in red) for a stack of three rigid pin gratings, which are aligned. The mode
trajectories intersect near a point identified from reflection and transmission properties of the stack as being $(\alpha_0,\beta)=(1.66451, 3.596951)$. The blue trajectory of minimum modulus is that which turns around; the red trajectory of minimum modulus is that for which the red lines concentrate around a bare region. 
Note that the resonant even mode in this case corresponds to  the left  eigenvector in equation (\ref{munsh13}).

\begin{figure}
\begin{center}
\includegraphics[width=8cm]{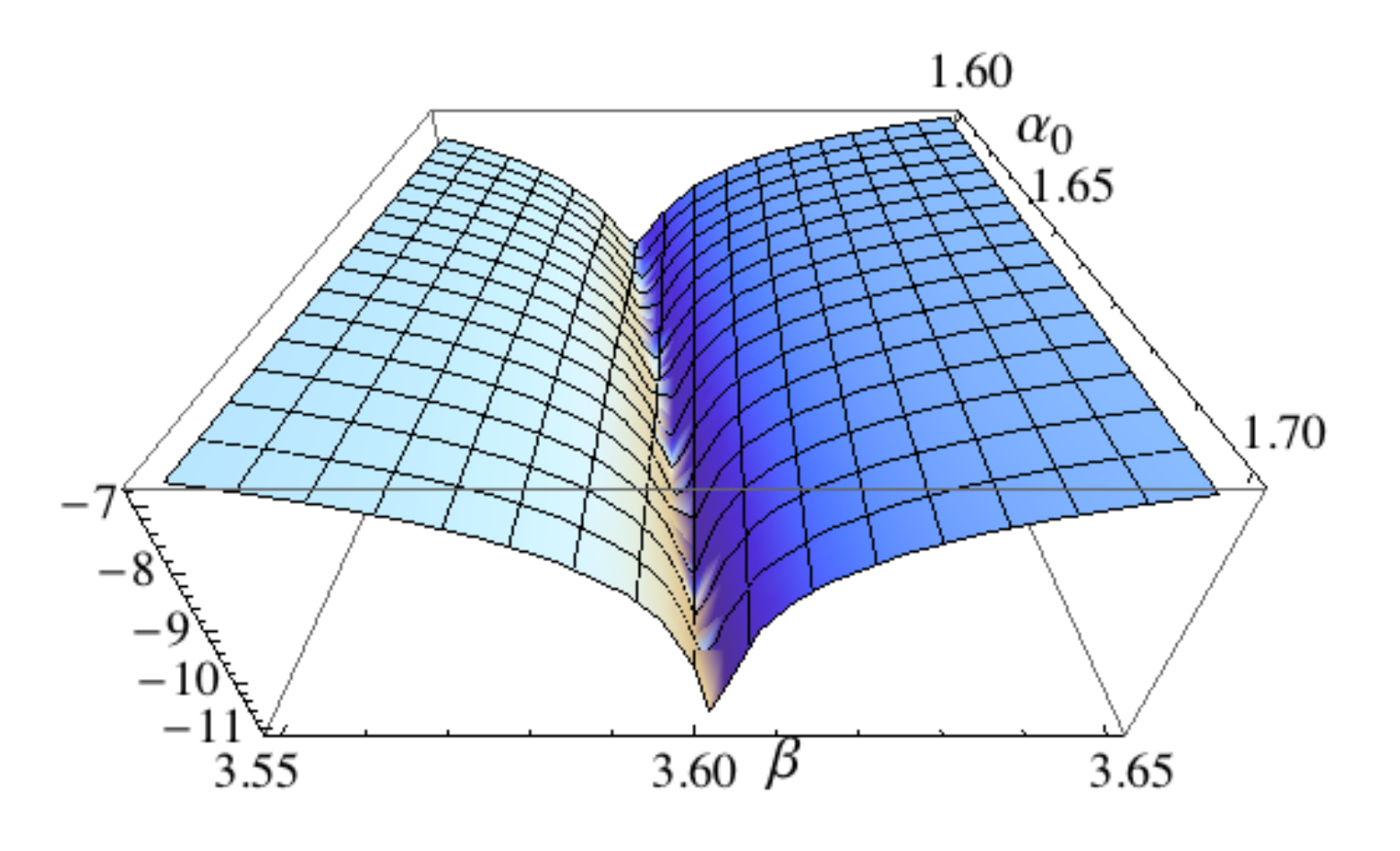}
\caption{\label{fig13exa}   Surface plot of the   modulus of $M_{11}-M_{13}$  for an aligned three grating stack of rigid pins,
as a function of $\alpha_0$ and $\beta$.}
\end{center}
\end{figure}

\begin{figure}
 \begin{center}
\includegraphics[width=6.8cm]{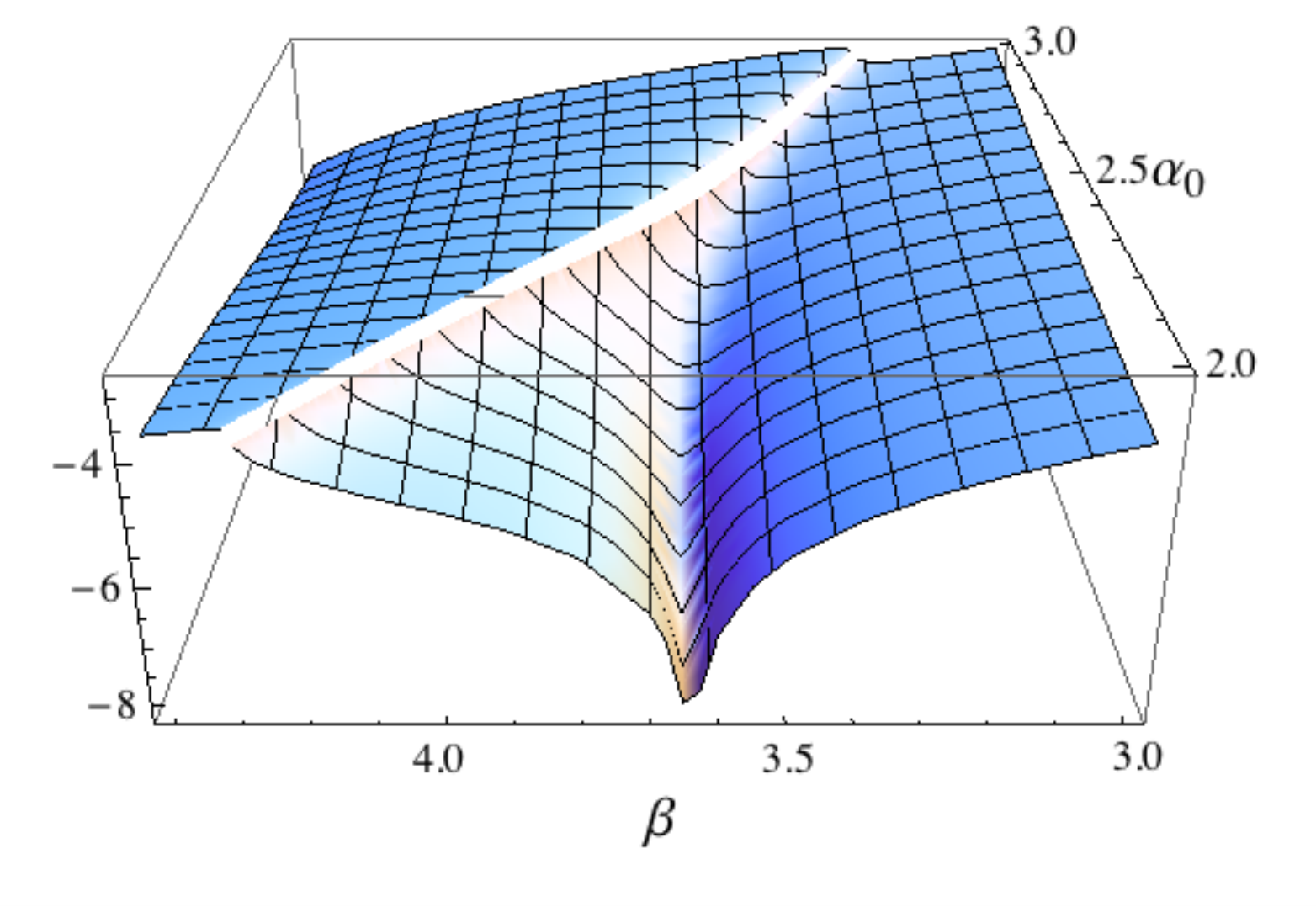}
\caption{\label{fig13exb}   Surface plot of the   modulus of $M_{11}-M_{13}$  for an aligned three grating stack of rigid pins,
as a function of $\alpha_0$ and $\beta$, giving the odd mode trajectory in the vicinity of the light line.}
\end{center}
\end{figure}

The surface plots in Figs.~\ref{fig13exa} and~\ref{fig13exb} show the values of $|M_{11}-M_{13}|$ as  functions of $\alpha_0$ and $\beta$. Deep valleys in the surface plots can be identified with regions where there is a well defined minimum and thus a well-identified mode. Fig.  \ref{fig13exa}  shows the region just after the crossing point of even and odd trajectories of Fig. \ref{fig13ex}, with the mode trajectory corresponding to the odd mode. In Fig.  \ref{fig13exb} we have moved to larger $\beta$ values, and are approaching the region of intersection of the light lines associated with grating orders $-1$ and $0$. Note that the odd mode modulus minimum  gradually gets weaker and weaker as we approach the intersection region. 

In order to study the even mode behaviour in more detail, it is valuable to split the three mode trajectories apart. Our first attempt to overcome the problem of modes washing out, arising from the reduction of the single grating's reflectance, used a projection method, based on the observation that the eigenvectors of each mode vary only weakly with $\beta$ and $\alpha_0$. For the 
normalised column eigenvectors identified in equations (\ref{munsh18}), (\ref{munsh12}) and (\ref{munsh13}) (denoted generically by ${\bf v}$) we form the
projection
\begin{equation}
p({\bf v})= {\bf v}^T {\bf M} {\bf v},
\label{munsh19}
\end{equation}
with the superscript $T$ denoting the transpose. This scalar just gives the estimate for the eigenvalue corresponding to $ {\bf v}$, assuming the accuracy of the eigenvector.

\begin{figure}
\subfigure[]{
\resizebox*{6.8cm}{!}{\includegraphics{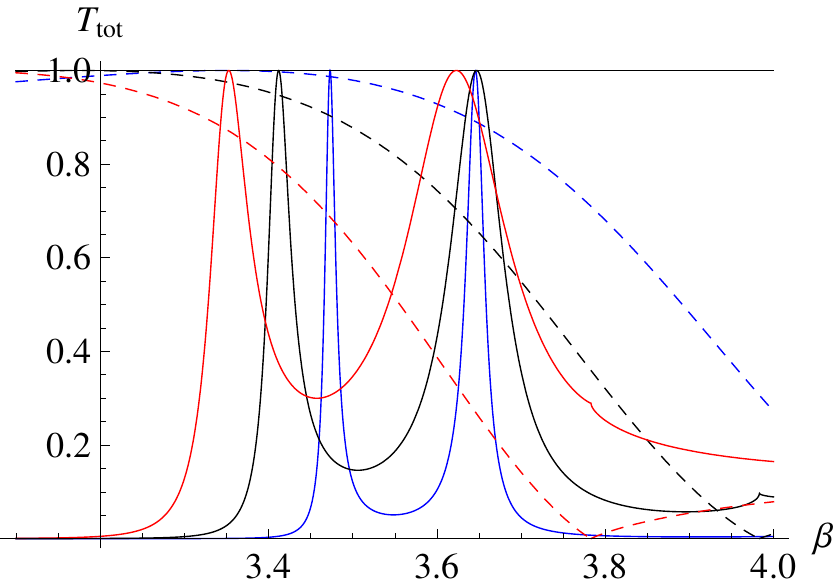}}}
\subfigure[]{
\resizebox*{6.8cm}{!}{\includegraphics{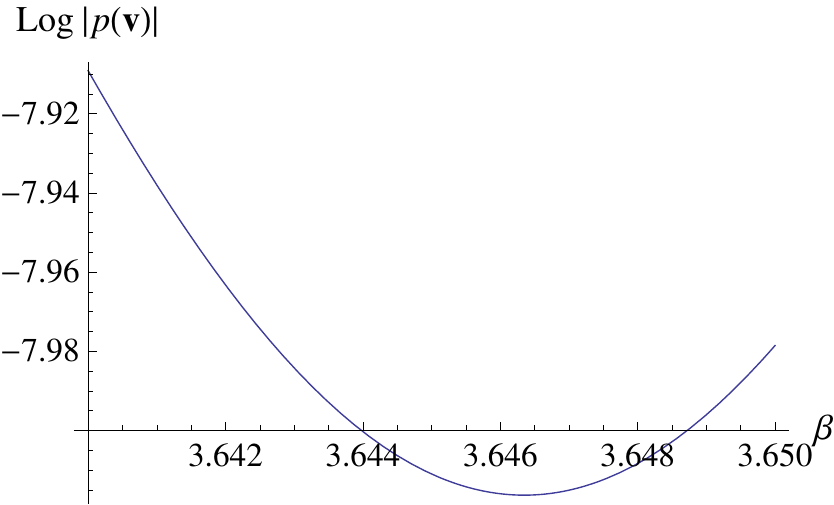}}}
\caption{\label{finesseplot}  (a) Normalised transmitted energy versus spectral parameter $\beta$ for a triplet of rigid pins with $\alpha_0 = 2.1$ (blue curve), $\alpha_0 = 2.3$ (black curve) and $\alpha_0 = 2.5$ (red curve). The corresponding curves for the reflectance of a single grating of rigid pins are dashed. (b) Log$|p({\bf v})|$ from equation~(\ref{munsh19}) for $\alpha_0 = 2.1$.}\end{figure}

We consider the case of $\alpha_0 = 2.1$ which supports two resonance modes, one even and one odd. The two modes are illustrated by the blue curve in Fig.~\ref{finesseplot}(a) and it is simple to classify them using equations~(\ref{munsh1})-(\ref{munsh13}). We determine the value of $\beta$ corresponding to the resonance on the right, which we expect to be odd. We substitute this value $\beta = 3.64581$ into~(\ref{munsh1}), obtaining the matrix $\BM$ and its corresponding eigenvalues and eigenvectors. 

Referring to equation~(\ref{oddc}), the third eigenvalue corresponds to a resonant mode, and its corresponding eigenvector matches up with the odd eigenvector defined by equation~(\ref{oddev}). We confirm that this mode is odd by using equation~(\ref{munsh19}) and plotting the logarithm of the absolute value of ${\bf v}^T {\bf M} {\bf v}$, where ${\bf v}$ is the odd mode taken from equation~(\ref{munsh12}) or~(\ref{munsh13}). This is shown in Fig.~\ref{finesseplot}(b) and the minimum occurs at around $\beta = 3.646$, the small discrepancy with the transmission resonance's value arising because of the slight difference in the eigenvectors.

\begin{figure}
\subfigure[]{
\resizebox*{5.8cm}{!}{\includegraphics{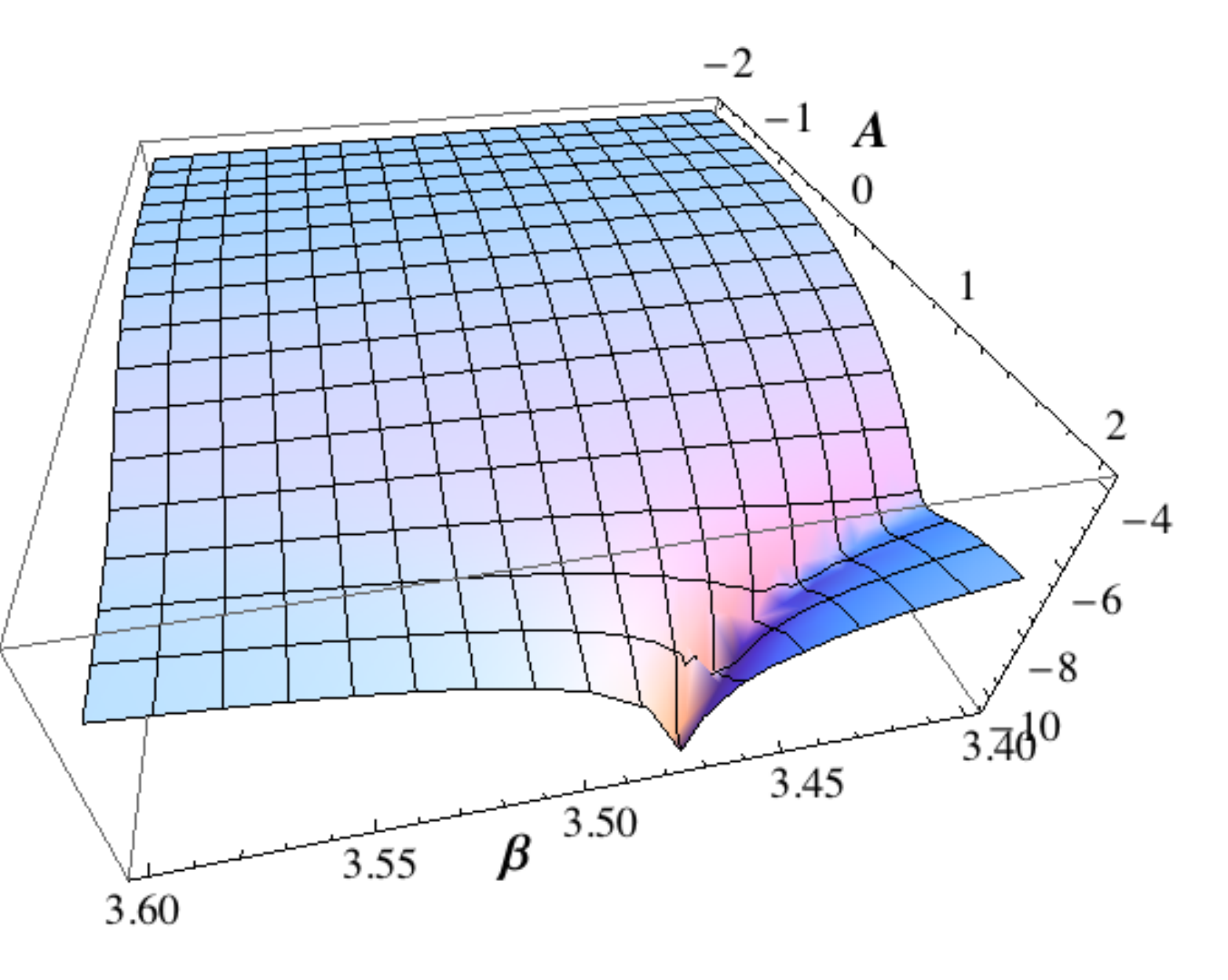}}}
\subfigure[]{\resizebox*{6.6cm}{!}
{\includegraphics{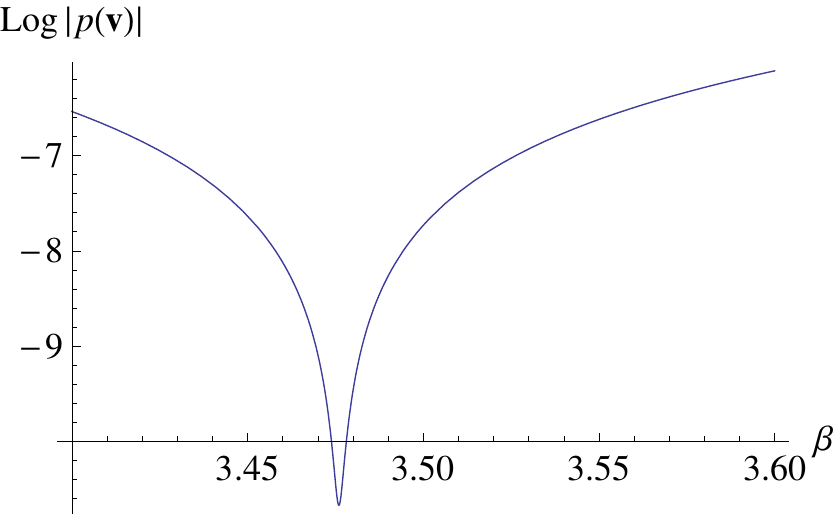}}}
\caption{\label{Aplot}   (a) Log$|p({{\bf v}_A})|$ with $\alpha_0 = 2.1$, $3.4 \le \beta \le 3.6$ and $-2 \le A \le 2$. (b)Log$|p({\bf v})|$ versus $\beta$ for ${\bf v}^T = 1/ \sqrt{6} (1, 2, 1)$. }\end{figure}

The other resonance for $\alpha_0 = 2.1$ possesses a higher $Q$ factor, as can be observed qualitatively in Fig.~\ref{finesseplot}(a), with $\beta = 3.473136$. We expect this mode to be of even type, with the characteristic equation~(\ref{evenc}). The corresponding eigenvector is one of the four types described by equations~(\ref{munsh12}) and~(\ref{munsh13}). We note that these even modes are of the form $${\bf v}_A \, = \, \frac{1}{\sqrt{A^2+2}}\left[\begin{array}{r} 1\\
A\\1\end{array}\right],$$
with $A$ in the range $-2 \le A \le 2$. We plot the surface for the logarithm of the absolute value of ${{\bf v}^T_A} {\bf M} {{\bf v}_A}$ with $-2 \le A \le 2$ and $\beta$ in a range including the transmission resonance's frequency $\beta = 3.473136$ in Fig.~\ref{Aplot}(a).
The result indicates that $A = 2$ defines the correct eigenvector and this can be confirmed by filling the matrix $M$  and determining the corresponding eigenvalues and eigenvectors.

We observe a zero eigenvalue for an eigenvector matching the first eigenvector of equation~(\ref{munsh13}). This indicates that the even mode arises from the equation 
\beq
M_{12}=-\sqrt {M_{11}(M_{11}+M_{13})/2},
\eequ{M12}
and if we apply equation~(\ref{munsh19}) to ${\bf v}^T = 1/ \sqrt{6} (1, 2, 1)$, we obtain the result matching the resonance frequency in Fig.~\ref{Aplot}(b).
We have therefore classified the two modes arising for this example of $\alpha_0$ to be odd, defined by the equation $M_{11} - M_{13} = 0$, and even defined by 
equation \eq{M12}.

\begin{figure}
\begin{center}
\includegraphics[width=10cm]{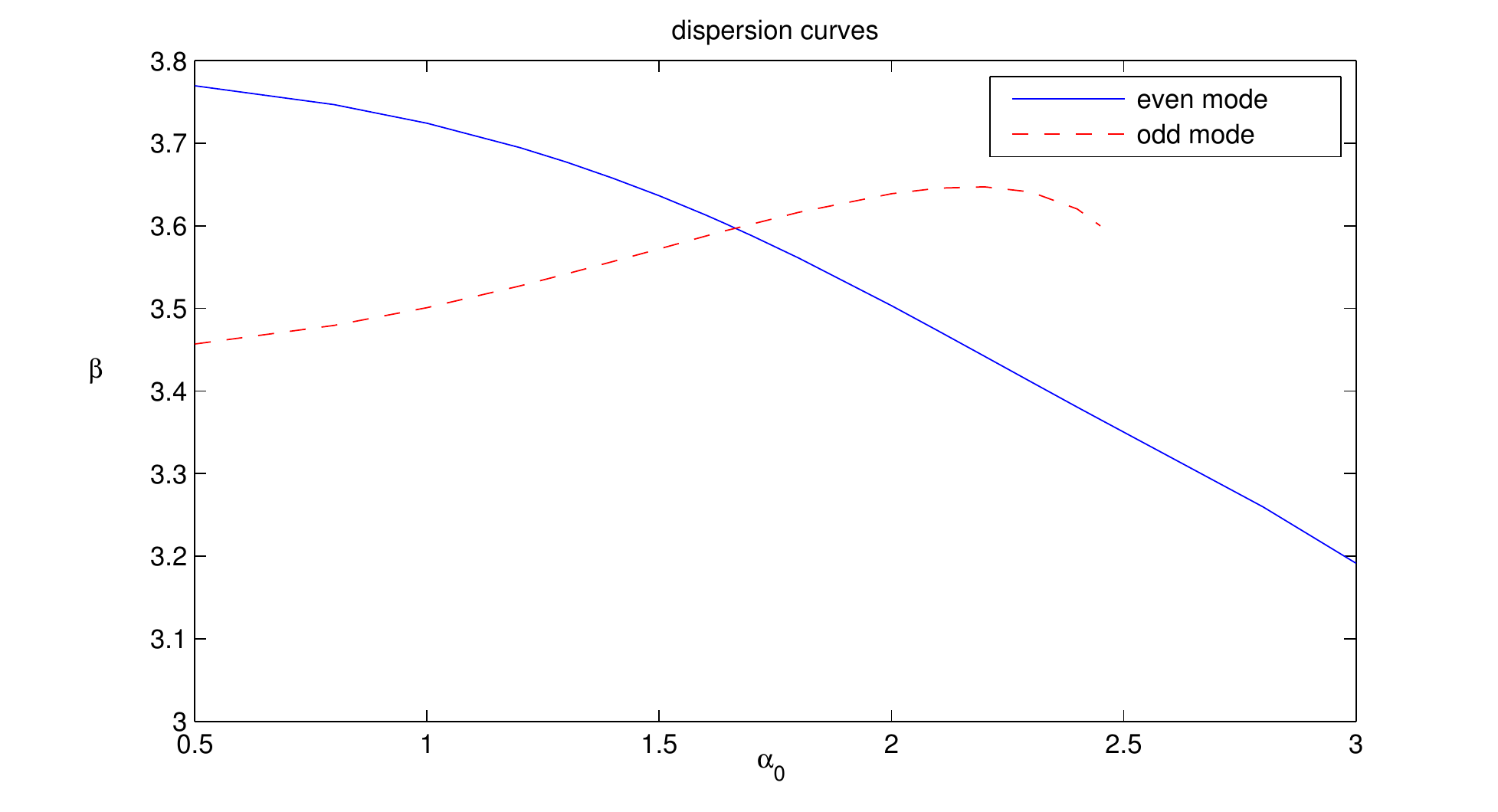}
\parbox{14cm}{\caption{\label{amdd04} Dispersion diagram for a waveguide consisting of an unshifted triplet with the horizontal axis representing $\alpha_0$ in the range $0.5 \le \alpha_0 \le 3$, and $\beta$ on the vertical axis in the range $3 \le \beta \le 3.8$. Produced using {\it Matlab}. The red curve represents the odd modes, and the blue curve, the even modes.}}
\end{center}
\end{figure}

 In general, the odd and even modes arise due to these two conditions and the corresponding dispersion curves are illustrated in Fig.~\ref{amdd04}. This initial method is useful for classifying the nature of modes and locating the possibility of EDIT effects where the modes coincide, but is only reliable for a narrow range of parameters. Referring to Fig.~\ref{amdd04}, the results for $1.2 \le \alpha_0 \le 2$ are well defined, but outside of this region, the mode trajectories begin to wash out, and the resonances become broader and lose resolution. This can be seen in Fig.~\ref{finesseplot}(a) where the transmission peaks are not aligned with the maxima for reflectance for a single grating, as well as in Figs.~\ref{fig13exb}, \ref{finesseplot}(b) and \ref{Aplot}(b) where the minima have relatively small magnitudes. We then developed a method that solves the two problems of mode symmetry classification and loss of high $Q$ simultaneously. This superior method involves optimizing the grating separation.

The evolution of mode behaviour from that characterised by narrow and deep minima to wide and ill-defined minima can be understood in terms of the reflectance of a single grating of pins.
If the reflectance of the single layer, $R_g$, is very high, then a pair or triplet of gratings can support very well confined guided waves between them. The corresponding resonant behaviour will then have a high quality factor, corresponding to a very localised region in $(\alpha_0,\beta)$ space. If we use the analogy with the classical Fabry-Perot interferometer, the transmittance of a pair
is given by
\begin{equation}
T_2=\frac{1}{1+F \sin^2 (\delta/2)}, ~~ {\rm where}~~F=\frac{4 R_g}{(1-R_g)^2}
\label{munsh20}
\end{equation}
is related to the finesse of the interferometer, and $\delta$ corresponds to the round-trip phase accumulated between the plates. In terms of resonant poles of the expression (\ref{munsh20}),
these will occur when $\delta=2 n\pi\pm i/\sqrt{F}$, so that the poles move away from the real $\beta$ axis in a way which scales as $1-R_g$. The bigger the finesse is, the narrower the resonance, and the closer it lies to the real axis. This model is useful qualitatively, but is not quantitatively accurate, since it applies to unstructured mirrors and the Helmholtz equation, not stacked gratings and the biharmonic equation.

\section{An Optimized Method to Steer Waveguide Modes}
The building block for the transmission problem and the waveguide is the single grating, which, provided its   reflectance is high, will then support the trapping of waves between multiple gratings with little leakage of energy. This idea allows the development of a simple procedure to construct  3-element systems that support extraordinarily high $Q$ resonances. For a given angle of incidence $\theta_i$, or Bloch parameter $\alpha_0$ for the analogous waveguide, we start by determining the value of $\beta= \beta_{\mbox{\small{g}}}$ for $R_{\mbox{\small{g}}} = |R_0|^2 = 1$ for the single grating. In the examples that follow we evaluate this reflectance to at least ten decimal places.

We then use $\beta_{\mbox{\small{g}}}$ and vary the separation parameter $\eta$ to find $\eta^*$ that determines the geometry for a pair of gratings to support an optimized trapped mode of odd up-down symmetry. This mode possesses an extremely high $Q$-factor and is that of the outer pair of gratings of a triplet, for which an even mode arises for the same geometry but for a different value of $\beta$. We finally use the additional parameter of lateral shift of the central grating $\xi$ to align the odd and even modes to the same value of $\beta$, thereby creating the EDIT effect for the chosen incidence angle. 

 \begin{figure}
 \subfigure[]{
\resizebox*{6.8cm}{!}{\includegraphics{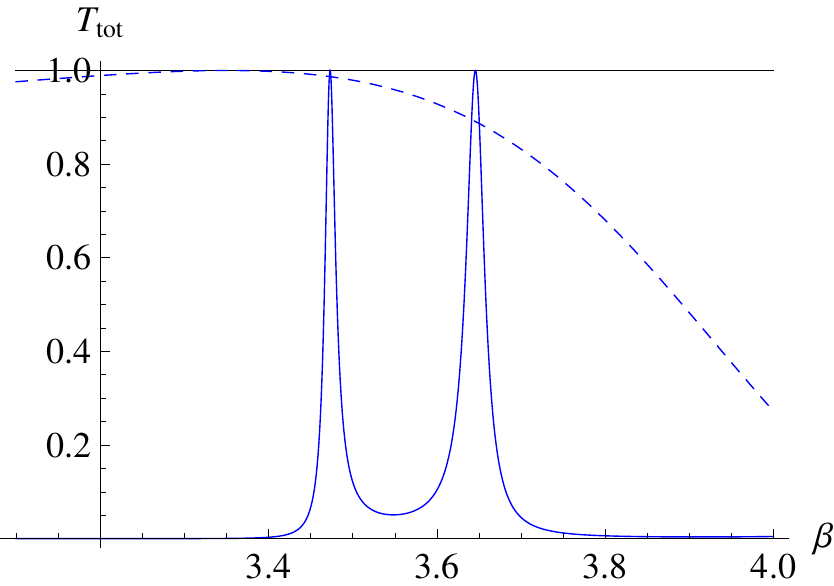}}}
\subfigure[]{\resizebox*{6.8cm}{!}
{\includegraphics{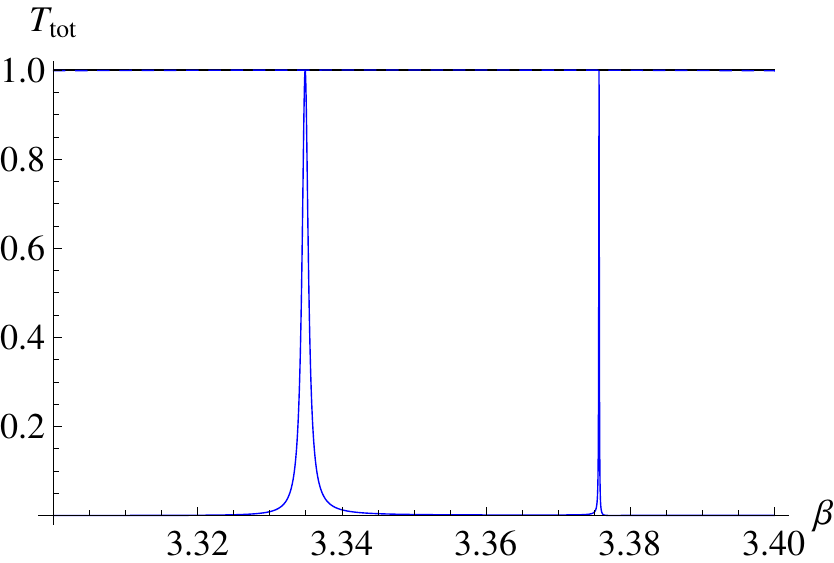}}}
\caption{\label{21opt} Normalised transmitted energy versus spectral parameter $\beta$ for a triplet of rigid pins with $\alpha_0  = 2.1$ in both cases: (a) $\eta = d$ ; (b) optimized resonances for a triplet with $\eta^* = 1.185266d$. }\end{figure}

An example of the second stage of this procedure is shown in Fig. \ref{21opt}. This shows (at left)  the transmittance spectrum in the vicinity of two transmission resonances, the lower $\beta$ one being of even symmetry and the upper of odd symmetry (see Fig. \ref{finesseplot}). With the initial vertical separation of the gratings being $\eta=d$, the resonances have low $Q$-factors, since the grating reflectance
is significantly below unity. At right, the transmittance spectrum for an optimised value of $\eta$ shows the resonances after they have been moved into the high reflectance region for the single grating:
both have much higher quality factors.

In optimising the grating separation, a useful guide is the model from optics of the dielectric slab waveguide studied in electromagnetic theory by, amongst others, Tien and Ulrich \cite{tien}.  In this model, we consider a slab of dielectric of thickness $h$, which is analogous to our grating separation $\eta$, and reflective index $n_f$, is situated between a cover region of index $n_c$ and a substrate region below of index $n_s$. 
Evanescent waves are present in the cover and substrate regions, but propagating plane waves that satisfy the Helmholtz equation arise in the dielectric region bordered by two boundaries $x = 0$ and $x = h$, on which consistency relations must be prescribed (electromagnetic boundary conditions). For non-trivial solutions, i.e. for modes to be trapped within the waveguide, a dispersion equation is derived from four equations as explained by Tien and Ulrich \cite{tien}. In our case, where our propagating waves also satisfy the Helmholtz equation, the situation is simplified by prescribing Dirichlet clamping conditions on $x = 0$ and $x = h$. We therefore obtain two equations corresponding to $x=h$ and $x=0$:
\begin{equation}
A e^{-i K_f h} \, + \, Be^{i K_f h} \, = \, 0, 
~ A  \, + \,B \, = \, 0, \label{ABK}
\end{equation}
where $K_f$ is the wave number.
It follows that for nontrivial solutions, $\sin{K_f h} = 0$ and therefore
$$
K_f \, = \, \frac{ \pi m}{h},$$
where $m$ is an integer. The analogue of $K_f$ in our case is $\chi_f = \beta \cos{\theta_f}$. Therefore for zeroth order propagating modes, $\chi_0 = \beta \cos{\theta_i}$ and we obtain the equation
\begin{equation}
h \, = \, \frac{\pi m}{\beta \cos{\theta_i}}=\frac{\pi m}{\sqrt{\beta^2-\alpha_0^2}}.
\label{slab}
\end{equation}

If we apply these equations to the data of Fig. \ref{finesseplot}, the optimized grating separation for a pair with $m = 1$ predicted by equation~(\ref{slab}) is $\eta^*=1.2031549$, while the value obtained by  locating numerically the value for which the pair transmittance is unity is $\eta^* = 1.185266d$, a difference of around 1.5\%. This magnitude of error arises from the differences between the simple but useful model of the dielectric slab waveguide in the case of the Helmholtz equation and the triplet of biharmonic grating structures.

In Table~\ref{tab:etao} we show computed values for optimal grating separation $\eta^*$ of order unity, and the corresponding value for $m$ in equation~(\ref{slab}).
\begin{table}
\begin{center}
\caption{Resonant frequencies $\beta_{\mbox{\small{g}}}$ and the corresponding optimal grating separation $\eta^*$ for pairs of unshifted gratings for various angles of incidence $\theta_i$.
In this table, $d = 1$.
}
\label{tab:etao}
\begin{tabular}{ccccc}
\hline\noalign{\smallskip}
$\theta_i$ & $\beta_{\mbox{\small{g}}}$ & $\alpha_0^{\mbox{\small{g}}}$ & $\eta^*$ & $m$  \\
\noalign{\smallskip}\hline\noalign{\smallskip}
$0^{\circ}$ & $4.456001$ & 0  & $0.6956042$ & 0.987 \\ 
$3^{\circ}$ &$4.438147$ & 0.232275 &0.698890 & 0.986   \\
$6^{\circ}$ & 4.387466  &  0.458615    &$0.708612$ & 0.984 \\ 
$9^{\circ}$ &$4.311191$ & 0.674419 &0.7244056 &  0.982  \\
$12^{\circ}$ & $4.217801$  & 0.87693    &$0.7458665$ & 0.979  \\  
  $15^{\circ}$&  4.11476 &  1.06498        &$0.77268$  & 0.978 \\ 
  $18^{\circ}$ & $4.007707$ & 1.23845    &$0.804674$  & 0.976\\  
  $21^{\circ}$ & $3.900536$  & 1.39783     &$0.841832$ & 0.976 \\ 
 $24^{\circ}$ &3.79580 & 1.54389    &$0.884279$ & 0.976\\ 
$27^{\circ}$&3.6950925  & 1.67754       &$0.932281$ & 0.979 \\  
 $30^{\circ}$ & $3.599363$  & 1.79968       &$0.98624$ & 0.977 \\ 
$33^{\circ}$& $3.509134$ & 1.91121       &$1.046715$ & 0.981\\
$36^{\circ}$ &3.424645   & 2.01296         &$1.114446$ & 0.983\\ 
$45^{\circ}$ &3.205694  & 2.26677       &$1.3723329$ &   0.990    \\ 
$60^{\circ}$ &2.94716   & 2.55232         &$2.12866291$ &0.998      \\ 
  \noalign{\smallskip}\hline
  \end{tabular}
    \end{center}
  \end{table}
We observe that there is good agreement between the separation equation and the data for $m = 1$. 
\begin{figure}
\begin{center}
\includegraphics[width=9cm]{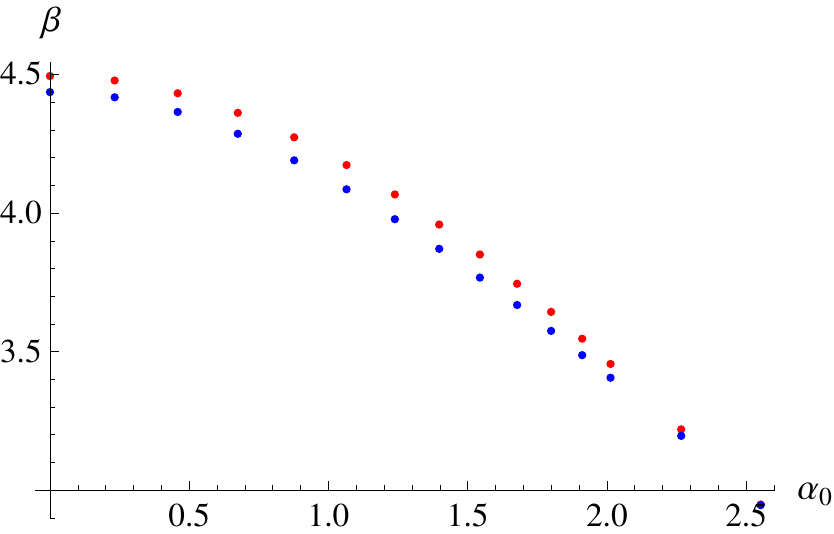}
\parbox{14cm}{\caption{\label{steering} $\beta$ versus $\alpha_0$ for odd (red) and even (blue) modes for optimised grating separation $\eta^*$ for unshifted triplets (Data from Table~\ref{tab:etao}).}}
\end{center}
\end{figure}
Using Table~\ref{tab:etao}, we can steer this mode by optimising the separation as $\theta_i$ is altered. This is illustrated in Fig.~\ref{steering}, where we show the even modes with the blue curve, and the odd modes with the red curve. We observe that by optimising the separation, we obtain a pair of modes that retain their relative position (i.e. higher $\beta$ for the odd mode) as well as their very high $Q$-factors. The modes approach one another very gradually, as can be seen by their proximity for $\alpha_0 = 2.55$, which corresponds to $\theta_i = 60^{\circ}$. 

\begin{figure}
\subfigure[]{
\resizebox*{6.8cm}{!}{\includegraphics{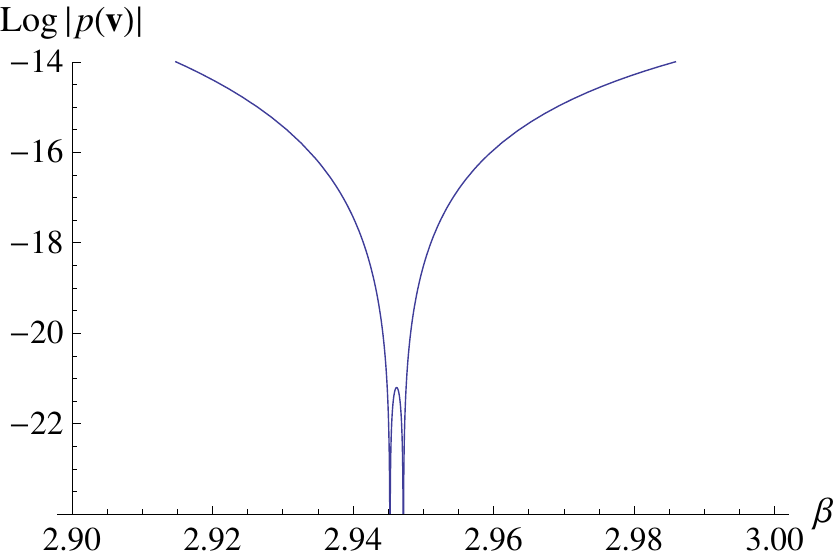}}}
\subfigure[]{\resizebox*{6.8cm}{!}
{\includegraphics{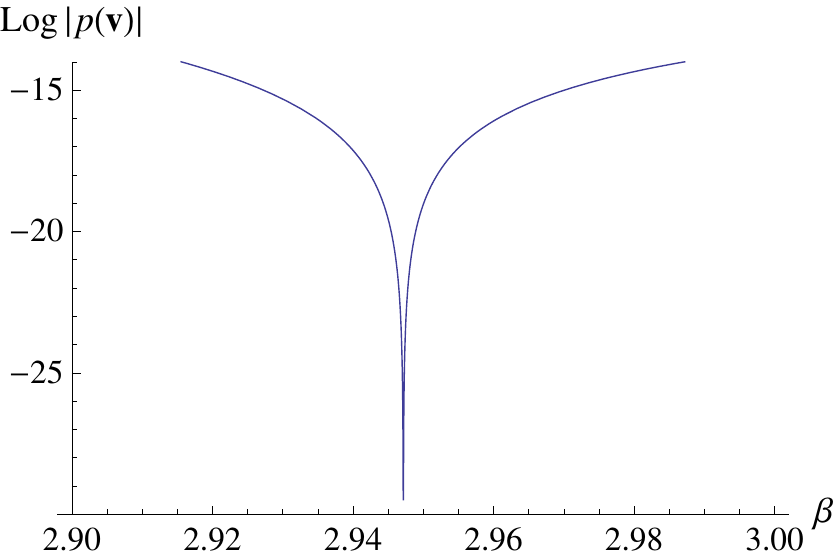}}}
\caption{\label{60deg}  The logarithm of the determinant of the matrix $\BM$ as a function of $\beta$ for $\theta_i = 60^{\circ}, \eta^* = 2.13196d$ (a) $\xi = 0$ (b) $\xi = 0.2476d$.}
\end{figure}

In the third stage of the EDIT steering procedure, the shift parameter $\xi$ is used to bring the two modes together. As a rough guideline, for each pair of odd and even modes plotted in Fig.~\ref{steering}, a shift in the vicinity of $\xi = 0.25200d$ is generally sufficient to produce EDIT effects. An example is shown in Fig.~\ref{60deg} for $\theta_i = 60^{\circ}$, where the actual optimal shift is
$\xi=0.2476 d$. A fast way to find the optimal shift for high quality resonances is through determinant plots of the type shown in Fig.~\ref{60deg}.
The outer pair in this example has grating separation $\eta = 4.262d$. 

\begin{figure}
\subfigure[]{
\resizebox*{6.8cm}{!}{\includegraphics{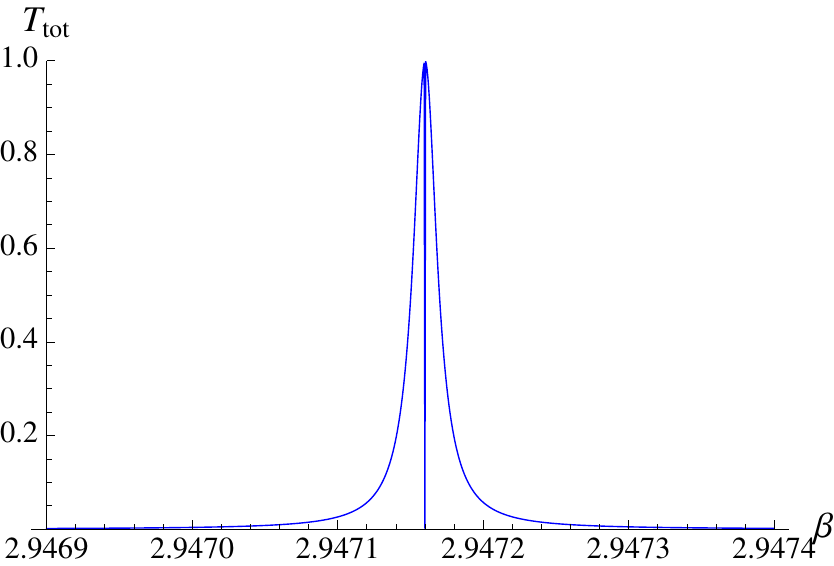}}}
\subfigure[]{\resizebox*{6.8cm}{!}
{\includegraphics{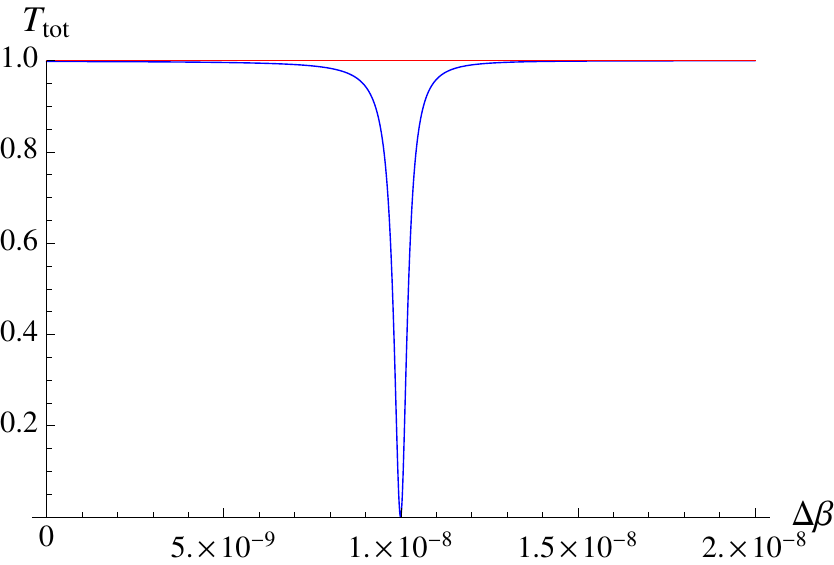}}}
\parbox{14cm}{\caption{\label{60edit} Optimized EDIT with $\theta_i = 60^{\circ}$, $\eta^* = 2.13196d$ and $\xi = 0.2476d$. In (b), we replace $\beta$ by 2.94715999 +  $\Delta \beta$, because the transmission minimum is so narrow.}}
\end{figure}

The coincidence of the two modes shown in Fig.~\ref{60deg}(b) for $\xi = 0.2476 d$ indicates the presence of the EDIT phenomenon for the transmission diagram, and this is given in Fig.~\ref{60edit}(a), with a detailed plot of the central minimum shown in Fig.~\ref{60edit} (b).
The peaks of transmittance either side of the EDIT dip reach $100 \%$, and the $Q$-factor of the notch is an extraordinarily high $6.5 \times 10^9$. The $Q$-factor of the outer pair of gratings is
around $1.50\times 10^5$,  a factor of forty thousand lower than that of the notch. This clear separation of scales in $\beta$ sensitivity means that the transmittance curves seem symmetric around the EDIT point, rather than having the asymmetric Fano shape \cite{fano} expected of a rapidly varying term superposed on a slowly varying one.

\begin{figure}
\subfigure[]{
\resizebox*{6.8cm}{!}{\includegraphics{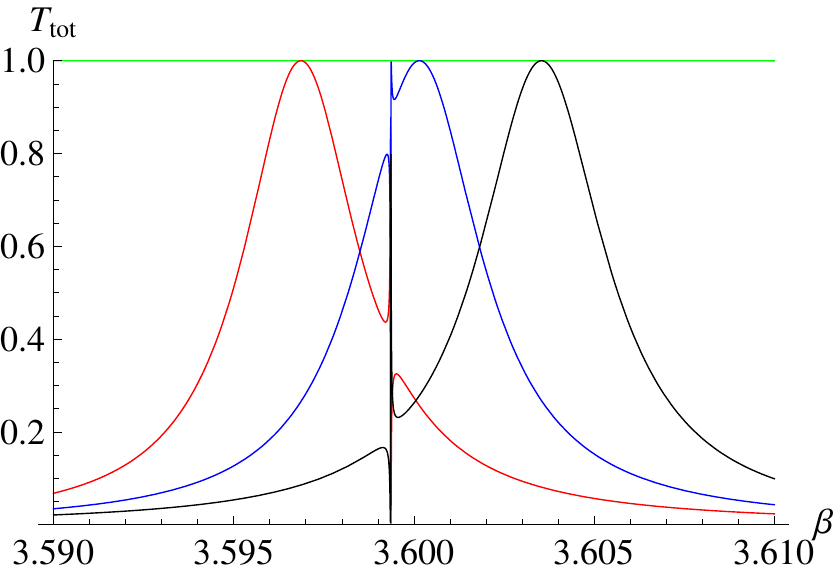}}}
\subfigure[]{\resizebox*{6.8cm}{!}
{\includegraphics{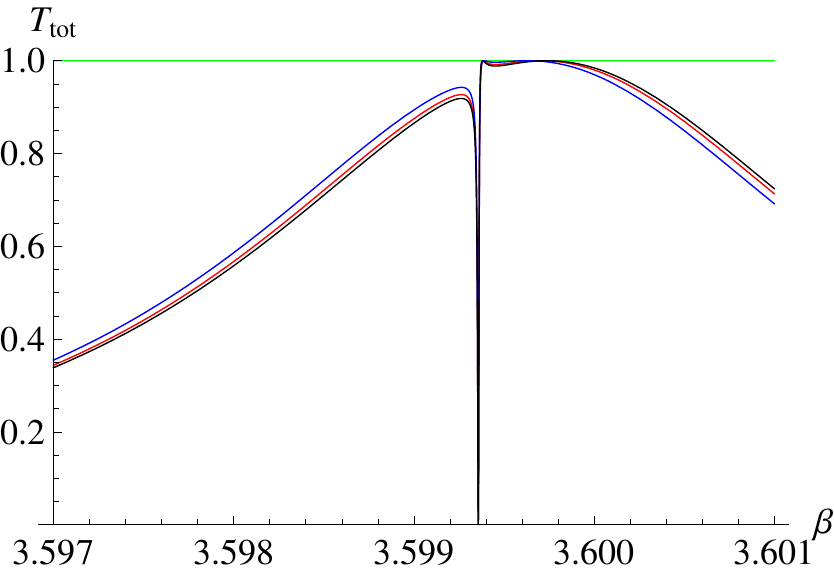}}}
\caption{\label{editfano}   Normalised transmitted energy versus spectral parameter $\beta$ for a triplet of rigid pins with $\theta_i = 30^{\circ}$ for various shifts of the central grating: (a) $\xi = 0.22d$ (red curve), $\xi = 0.23d$ (blue curve) and $\xi = 0.24d$ (black curve) (b) $\xi = 0.2282d$ (red curve), $\xi = 0.2283d$ (blue curve) and $\xi = 0.2284d$ (black curve).}
\end{figure}

In Figure~\ref{editfano} we show the variety of filtering curves which may arise for variations of the shift parameter $\xi$ in the neighbourhood of the EDIT value. The very sharp rise in transmittance from
zero to unity which occurs when the shift is just below the EDIT value is the result of the even mode resonance occurring  just before the odd mode resonance. This sharp rise might be useful in some filtering applications.

\begin{figure}
\begin{center}
\includegraphics[width=3.0in]{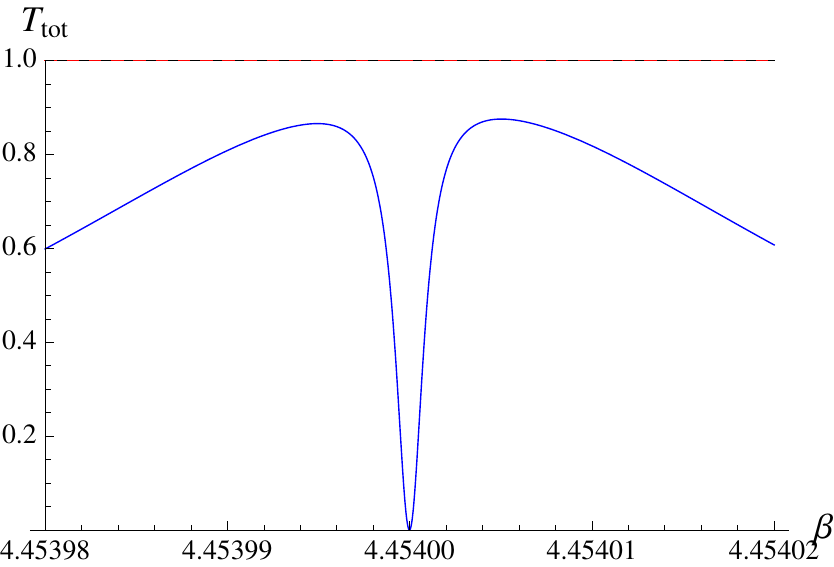}~
\caption{\label{1deg}   Normalised transmitted energy versus spectral parameter $\beta$ for a triplet of rigid pins with $\theta_i = 1^{\circ}$ for the optimized grating separation $\eta = 0.705679367d$ and the EDIT shift of the central grating $\xi = 0.165868d$.} 
\end{center}
\end{figure}

As the angle of incidence approaches zero, resonances due to the even modes become increasingly sharp. This makes the task of locating and optimizing the EDIT phenomenon increasingly difficult. The transmission diagram shown in Fig.~\ref{1deg} is the best we have been able to obtain for angle of incidence $\theta_i = 1^{\circ}$, and we have so far been unsuccessful in detecting the EDIT minimum for exactly normal incidence.

We conclude by stressing that the procedure we have demonstrated for steering the angle and frequency of the EDIT phenomenon makes it much more flexible than its analogue of EIT in atomic physics, for which the frequency is fixed by the occurrence of a set of appropriate energy levels. It may be that a similar procedure may be adapted to solutions of the Helmholtz equation, and other wave-bearing equations.

{\bf Acknowledgements.}
S.G.H. gratefully acknowledges the financial support of the Duncan Norman Charitable Trust through the Duncan Norman Research Scholarship. N.V.M. and R.C.M. acknowledge the financial support of the European Community's Seven Framework Programme under the contract number PIAPP-GA-284544-PARM-2. R.C.M. also acknowledges support from the Australian Research Council through its Discovery Grants Scheme.

\label{lastpage}

\end{document}

%% file: macro.tex
\newcommand\eq[1] {(\ref{#1})}

\newcommand{\beqa}{\begin{eqnarray}}
\newcommand{\eeqa}[1]{\label{#1}\end{eqnarray}}
\newcommand{\bequ}{\begin{equation}}
\newcommand{\eequ}[1]{\label{#1}\end{equation}}
\def\Ba{{\bf a}}

\def\BA{{\bf A}}

\def\BM{{\bf M}}

\def\BU{{\bf U}}

\newcommand{\beq}{\begin{equation}}
\newcommand{\eeq}{\end{equation}}
\newcommand{\overliner}{\begin{eqnarray}}
\newcommand{\earr}{\end{eqnarray}}
\newcommand{\beqn}{\begin{equation*}}
\newcommand{\eeqn}{\end{equation*}}
\newcommand{\overlinern}{\begin{eqnarray*}}
\newcommand{\earrn}{\end{eqnarray*}}